\documentclass[letterpaper, 10 pt, conference]{ieeeconf}

\IEEEoverridecommandlockouts   

\usepackage[utf8]{inputenc}
\usepackage{amsmath}
\usepackage{mathtools}
\usepackage{color}
\usepackage{graphicx}
\usepackage{caption}
\usepackage{subcaption}
\usepackage[ruled,vlined]{algorithm2e}
\usepackage{amssymb,amsmath,amsfonts}

\newtheorem{remark}{Remark}%[section]

\title{Utilizing synchronization to partition power networks into microgrids}
\author{Ricardo Cardona-Rivera$^*$, Francesco Lo Iudice$^*$ $^\dagger $, Antonio Grotta, Marco Coraggio, Mario di Bernardo
\thanks{$^*$ These authors contributed equally. $^\dagger$Corresponding author; e-mail: francesco.loiudice2@unina.it. 
 \newline All authors are with the Department of Electrical Engineering and Information Technology, University of Naples Federico II, 80125, Naples, Italy. Mario di Bernardo is also with the Department of Engineering Mathematics, University of Bristol, Bristol, UK.}}
 \date{February 2021}
\begin{document}
\maketitle
\begin{abstract}
The problem of partitioning a power grid into a set of microgrids, or islands, is of interest for both the design of future smart grids, and as a last resort to restore power dispatchment in sections of a grid affected by an extreme failure. In the literature this problem is usually solved by turning it into a combinatorial optimization problem, often solved through generic heruristic methods such as Genetic Algorithms or Tabu Search \cite{haddadian2017multi,arefifar2013optimum}. 
In this paper, we take a different route and obtain the grid partition by exploiting the synchronization dynamics of a cyberlayer of Kuramoto oscillators, each parameterized as a rough approximation of the dynamics of the grid's node it corresponds to. We present first a centralised algorithm and then a decentralised strategy. In the former, nodes are aggregated based on their internode synchronization times while in the latter they exploit synchronization of the oscillators in the cyber layer to self-organise into islands. Our preliminary results show that the heuristic synchronization based algorithms do converge towards partitions that are comparable to those obtained via other more cumbersome and computationally expensive optimization-based methods. 
\end{abstract}

%\section{Outline}

%\begin{itemize}
%    \item motivation: the partitioning problem arises in both planning and extreme fault recovery situations;
 %   \item motivation b: network partitioning as an (inefficient) optimization problem
%    \item problem formulation 
 %   \item problem solution: exploiting synchronization of Kuramoto oscillators for network partitioning
 %   \begin{itemize}
  %      \item a centralized solution
   %     \item a decentralized solution
    %\end{itemize}
%    \item Numerical Validation
%    \begin{itemize}
%        \item the IEEE 9-node test case (or the 14 node test case): we could use this test case to explain in detail the application of our two algorithms. One three panel figure dedicated to this example with the test case and the two partitions (that obtained through the centralized algorithm and that obtained through the decentralized algorithm);
%        \item the IEEE 118 test-case: this could be an example on which we give a quantitative assessment of the performance of our algorithms both in terms of structural differences with respect to existing partitions of the same test case (that by Emanuele), and in terms of the metrics we found in the literature. Again at least a three panel figure here, we need two panels for the metrics, and maybe two for the partitions (we could avoid portraying the original network in this case). Just a news, the centralized solution is very close to that of Emanuele.
%    \end{itemize}
%    \item Conclusions
%\end{itemize}

\section{Introduction}
Power networks have become a topic of pressing interest for researchers in the Control Engineering community because of the novel control challenges introduced by the penetration of renewable and distributed generation throughout their hierarchical control architecture, e.g., \cite{dorfler2015breaking,bidram2012hierarchical,frasca2015distributed}. The need of ensuring that the frequency of the output AC signals of Voltage Source Converters (VSCs) is coherent with that of the remainder of the grid \cite{rocabert2012control,tayyebi2020frequency,arghir2018grid} has generated new primary control problems. Moreover, in the secondary and tertiary layers \cite{milano2018foundations,dorfler2019distributed}, it has become even more critical to solve the frequency regulation problem and to optimize the non-renewable power generation share because of the inherent uncontrollable and stochastic nature of renewable power sources \cite{lalor2005frequency,bevrani2010renewable,ulbig2014impact}. A possible solution is to identify sections of the grid that can, if needed, isolate and operate independently from it. In such a distributed generation framework, a power grid can then be designed so as to allow under certain circumstances some of its sections to isolate into a \emph{network of microgrids} or \emph{multi-microgids}  \cite{wang2014coordinated,nikmehr2015optimal,haddadian2017multi}. Such an operation can also be used to guarantee a better degree of resilience to extreme events such as cascading failures so that power can be dispatched at least in some portions of the failed grid (the so called islanding problem described, for instance, in \cite{Fan2021Controlled,pahwa2013optimal,sun2003splitting,adibi2006power,fernandez2018intentional}).

Partitioning a grid into a set of microgrids or islands is usually modelled as a combinatorial problem \cite{arefifar2012supply} and sometimes turned into a graph optimization problem. Often, solving these problems numerically is cumbersome or inefficient so that heuristic strategies are frequently used to seek a suboptimal solution \cite{haddadian2017multi,pahwa2013optimal,arefifar2012supply,arefifar2013optimum,arefifar2014dg,moghateli2020multi,hasanvand2017new,mohammadi2014scenario}.

In this letter, we propose to use the synchronization dynamics among the nodes of a \emph{cyberlayer} associated to the physical grid in order to obtain a heuristic solution to the grid partitioning problem. The cyberlayer is obtained by associating to each node in the grid the equations of a first-order Kuramoto oscillator parameterized so as to provide a rough approximation of the dynamics of the physical network node it corresponds to. Nodes in the cyberlayer are then interconnected as those in the physical grid and, starting from random initial conditions, their synchronization dynamics is observed. We propose two different approaches to partition the grid using synchronization. Firstly, we present a \emph{centralised strategy} where nodes are aggregated on the basis of how quickly they synchronize among each other on the cyberlayer. Secondly, we present a \emph{decentralised} alternative where the load nodes exploit synchronization of the nodes in the cyberlayer to estimate the power imbalance in neighboring islands. These estimates are then used by each node to decide whether to attach or not to an island depending on whether the latter has the ability of meeting its power requirement. We wish to emphasize that while the centralised strategy could be a valuable tool to construct a network of microgrids by design, the decentralised approach could be deployed online in response to cascading failures.

To validate the strategies, we apply both to the benchmark IEEE 118 network, comparing the viable partitions we obtain to others suggested in previous papers in the literature. 

\section{Preliminaries and Problem Statement}
 We model a power grid as an undirected connected graph $\mathcal{G}(\mathcal{V},\mathcal{E})$. The set of $n$ grid nodes $\mathcal{V}$ consists of two subsets, that of the $n_g$ generators $\mathcal{V}^{\text{gen}}$ and that of the $n_l$ loads $\mathcal{V}^{\text{load}}$. An edge $(i,j)$ is in the set $\mathcal{E}$ iff there exists a transmission line connecting nodes $i$ and $j$. The adjacency matrix $A$ of the graph $\mathcal{G}$ is the symmetric matrix whose $ij$-th element $a_{ij} \neq 0$ if $(i,j)\in \mathcal{E}$ and $a_{ij} = 0$ otherwise. A graph $\mathcal{G}_i(\mathcal{M}_i,\mathcal{E}_i)$ is a subgraph of $\mathcal{G}$ if $\mathcal{M}_i \subseteq \mathcal{V}$, and $\mathcal{E}_i \subseteq \mathcal{E}$. Moreover, we say that the grid described by the graph $\mathcal{G}$ is {\em partitioned} into $n_{\mu}$ microgrids (or islands) described by the subgraphs $\mathcal{G}_i \subset \mathcal{G}$, $i=1,\dots,n_{\mu}$ if $\cup_{i=1}^{n_{\mu}}\mathcal{M}_i = \mathcal{V}$ and $\cup_{i=1}^{n_{\mu}}\mathcal{E}_i = \mathcal{E}\setminus \lbrace (l,k): l\in \mathcal{M}_i, \ k\in \mathcal{M}_j, \ i\neq j   \rbrace  $. We denote by $\mathcal{N}_j$ the set of neighbours of node $j$ and, given a set of nodes $\mathcal{M}_i$, by $\mathcal{N}(\mathcal{M}_i)$ the set of all nodes which are neighbours of the nodes in $\mathcal{M}_i$, i.e. $\mathcal{N}(\mathcal{M}_i) :=\lbrace j:\exists l \in \mathcal{M}_i:(j,l)\in \mathcal{E} \rbrace$. Finally, given a set $\mathcal{M}$ we denote its cardinality by $|\mathcal{M}|$.

The problem of partitioning a grid in $n_{\mu}$ microgrids is usually stated as that of finding the node sets $\mathcal{M}_1,\dots, \mathcal{M}_{n_{\mu}}$ of the set $\mathcal{V}$ such that
\begin{align}
    \underset{\mathcal{M}_1,\dots, \mathcal{M}_{n_{\mu}}}{\min} J(\mathcal{M}_1,\dots, \mathcal{M}_{n_{\mu}})\\
    \cup_{i}\mathcal{M}_i = \mathcal{V}\\
    \mathcal{M}_j\cap_{i}\mathcal{M}_i = \emptyset \ \forall i\neq j\\
    \mathcal{G}_i \ \mathrm{is \  connected} \ \forall i.
    \end{align}
with $J$ being an appropriated selected cost function; see Table 1 in \cite{haddadian2017multi} for an overview of different choices of $J$ made in the literature, or \cite{Fan2021Controlled,pahwa2013optimal,sun2003splitting,adibi2006power,fernandez2018intentional} for further examples. As anticipated, here we will take a different heuristic approach based on synchronization to obtain a viable grid partition. We provide in section \ref{sec:numerical} a comparison between our results and some benchmarks which were previously obtained in the literature as a solution of the optimization problem described above.

\section{A synchronization based approach}\label{sec:synch_for_sol}

%\subsection{Microgrid initialisation}\label{subsec:init}
The first step of our approach is to start from an initial arbitrary partition of the grid into $n_{\mu}$ microgrids, say $\mathcal{M}_1, \dots, \mathcal{M}_{n_{\mu}}$, by assigning to each of them an arbitrary subset of the $n_g$ generators in the network. For instance, an initial partition could be obtained by considering the largest generators in the grid such as synchronous machines or large VSCs so as to ensure each microgrid is at least partially independent.

If any graph, say $\mathcal{G}_i(\mathcal{M}_i,\mathcal{E}_i)$, obtained from this initialisation is not connected, we add to $\mathcal{M}_i$ the nodes crossed by the shortest paths connecting the nodes in $\mathcal{M}_i$.
\begin{remark}
When the aim of the partition is to solve the islanding problem in response to an extreme failure, a fundamental constraint in the electrical engineering literature is that this initial partition be determined on the basis of the sets of coherent generators \cite{sastry1981coherency,lin2018wams,wei2013multi}. 
%When instead the aim is that of designing a multi-microgrid network, then this initial partition could be the result of preliminary design considerations. 
\end{remark}

\subsection{A centralised partitioning strategy}

In our centralised approach, we associate to the physical grid a cyberlayer where each node corresponding to a node in $\mathcal{V}$ is a Kuramoto oscillator. The natural frequency of the $i$-th oscillator is chosen to be the power absorbed or generated by node $i$ in the physical grid. The oscillators are coupled in the cyberlayer via a graph sharing the same structure as that of the underlying grid, $\mathcal{G}(\mathcal{V},\mathcal{E})$, whose edges are weighed with the susceptances $B_{ij}$ of the transmission lines of the grid. Hence, the dynamics of the cyberlayer can be described as
\begin{equation}\label{eq:Kur_approx}
        \dot \theta_i = P_i + \sum_{j\in \mathcal{V}}B_{ij}\sin(\theta_j-\theta_i),
\end{equation}
where $\theta_i$ is the phase of the $i$-th oscillator capturing a rough approximation of the nodes' dynamics.

For each pair (i,j) of nodes such that $B_{ij}\neq 0$, we define the {\em internode synchronization time} as 
\begin{equation}
    t_{ij}: = \lbrace t^*: \rho_{ij}(t)>0.99 \ \forall t\geq t^* \rbrace,
\end{equation}
where 
\begin{equation}\label{eq:ord_param}
    \rho_{ij}(t) = \frac{1}{N_s}\sum_{s=1}^{N_s}\cos(\theta_{i,s}(t)-\theta_{j,s}(t)) \ \in [0,1].
\end{equation}
is obtained by running $N_s$ simulations of the dynamics of the cyber layer starting from a set of initial conditions randomly extracted from a uniform distribution in the interval $]-\pi/2, \pi/2]$; $\theta_{i,s}(t)$ being the phase of node $i$ at time $t$ in the $s$-th simulation. We define
\begin{equation}
    \mathcal{T}:=\lbrace t_{ij} : (i,j) \in \mathcal{E}\rbrace
\end{equation}
as the set of all internode synchronization times obtained from the set of $N_s$ numerical simulations. To obtain the power grid partition we then apply Algorithm 1.   
\begin{algorithm}[t]
\SetKwInOut{Input}{input}\SetKwInOut{Output}{output}
 \Input{Power network; set of intial microgrids $\mathcal{M} = \cup_{i=1}^{n_{\mu}}\mathcal{M}_i$;
 set of synchronization times $\mathcal{T}$.}
 \Output{A partition of the power network.}
 \BlankLine
 %initialization\;
 \While{$\mathcal{M} \neq \mathcal{V}$}{
 %$t_\text{prev} \leftarrow \infty$\;
 compute $\Delta P(\mathcal{M}_i):= \sum_{j\in\mathcal{M}_i} P_j \ \forall i$\;
 $i^* \leftarrow \text{argmax}_{i = 1,\dots,n_{\mu}} \Delta P(\mathcal{M}_i )$\;
        $l^* \leftarrow \text{argmin}_{l \in \mathcal{N}(\mathcal{M}_{i^*}), \ l\notin \mathcal{M}} \lbrace t_{jl} : j\in \mathcal{M}_{i^*} \rbrace$\;
        $\mathcal{M}_{i^*} \leftarrow \{\mathcal{M}_{i^*},l^*\}$\;
 }
\caption{Centralized partitioning}
\label{alg:Centralized_Alg}
\end{algorithm}

\subsection{A decentralised partitioning strategy}
\label{sec:decentralised_partitioning}

\begin{figure}[t]
  \centering
    \includegraphics[scale=0.8]{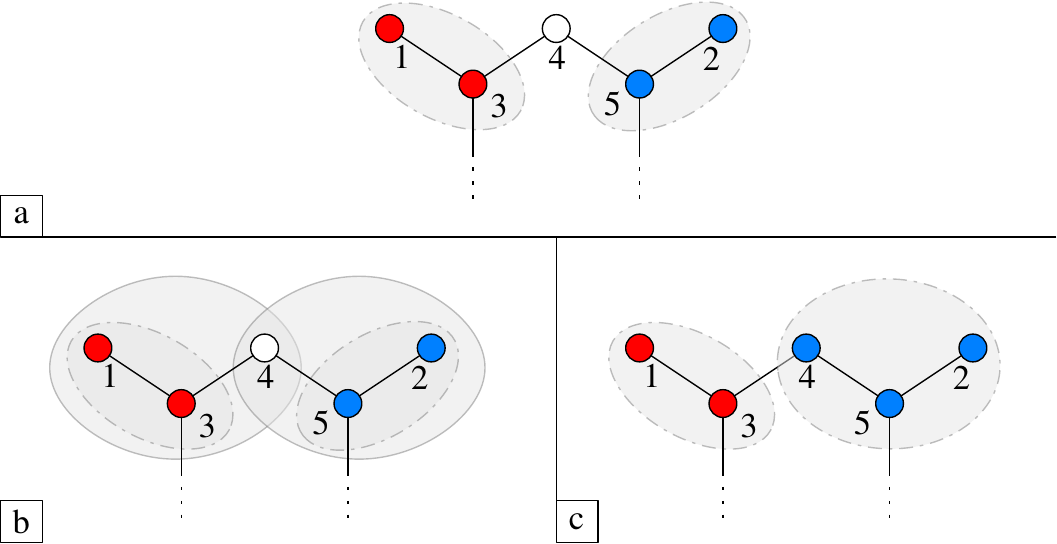}
\caption{Example of the decentralized partitioning algorithm described in Section \ref{sec:decentralised_partitioning}. 
Panel a: starting islands are fundamental layers. 
Panel b: additional cyber layers are created. %and simulations of Kuramoto dynamics \eqref{eq:Kur_cyb} are run over $\mathcal{G}_1$, $\mathcal{G}_2$, $\mathcal{G}_1^3$ and $\mathcal{G}_2^5$.
%Panel c: Nodes 4 attaches to islands 1 and 2, respectively; next the simulations of Kuramoto dynamics are run over the updated cyber layers $\mathcal{G}_1$, $\mathcal{G}_2$, $\mathcal{G}_1^4$ and $\mathcal{G}_2^4$.
Panel c: Letting 4 be a load, it is attached to the blue island, as it has a larger(positive) power balance. 
The Fundamental layers are then updated.}
\label{fig:representation_decentralized}
\end{figure}

We next provide a decentralised strategy where the network nodes self-organize to form a set of islands partitioning the grid. 
Once the initialisation is performed as described earlier on, each of the islands is associated to a cyberlayer of Kuramoto oscillators sharing its same structure. 
Specifically, let $\mathcal{G}_l(\mathcal{M}_l,\mathcal{E}_l)$ be the graph describing the structure of the $l$-th island (see the sketch diagram in Figure \ref{fig:representation_decentralized}a). 
Then, the dynamics of its corresponding cyberlayer is chosen to be  
\begin{subequations}\label{eq:Kur_cyb}
\begin{equation}
    \dot \theta_i = P_i + \sum_{j\in \mathcal{M}_l}B_{ij}\sin(\theta_j-\theta_i)
\end{equation}
\begin{equation}
\theta_i(0) = 0    
\end{equation}
\end{subequations}
with $P_i$ and $B_{ij}$ defined as before.

Note that, from \cite{DoBu:14}, it can be shown that when (\ref{eq:Kur_cyb}) reaches synchronization, the frequency $\bar{\omega}_l$ of its synchronous solution can be computed as a function of the power imbalance of microgrid $l$ defined as
\begin{equation}\label{eq:excess}
    P(l):=\sum_{j\in\mathcal{M}_l}P_j
\end{equation}
In particular, from \cite{DoBu:14}, we have
\begin{equation}\label{eq:synch_freq_i}
    \bar \omega_l = \frac{P(l)}{|\mathcal{M}_l|}.
\end{equation}

Then, a node in the grid, say the $i$-th node, that is not already assigned to an existing microgrid, can run the following local algorithm to decide to which one of its neighboring $L_i$ microgrids it should get attached to. To better illustrate the algorithm we refer to the example reported in Figure \ref{fig:representation_decentralized} where the network needs to be partitioned into $n_\mu=2$ microgrids.

\begin{enumerate}
\item The first step is to initialize the algorithm by choosing two arbitrary islands. In the example depicted in Figure \ref{fig:representation_decentralized}(a) we set the node set of island $1$ to $\mathcal{M}_1(0) = \{1,3\}$, and the node set of island $2$ to $\mathcal{M}_2(0) = \{2,5\}$. 
\item Each island runs its associated cyberlayer, say $\Gamma_1$ and $\Gamma_2$, each of the form (\ref{eq:Kur_cyb}) until reaching synchronization on a solution with frequency $\bar \omega_{1}$ for $\Gamma_1$ and $\bar \omega_{2}$ for $\Gamma_2$.

\item Node 4, which is not part of any of the two initial islands, receives from nodes 3 and 5 the synchronization frequencies of the two islands it is neighboring with. Moreover, it runs locally the simulation of two additional cyberlayers, $\Gamma_{1}^4$ whose node set is $\mathcal{M}_1^4 = \{1,3,4\}$  and $\Gamma_{2}^4$ whose node set is $\mathcal{M}_2^4 = \{2,5,4\}$ (see Figure \ref{fig:representation_decentralized}(b)). Hence, such cyberlayers consist of those corresponding to the original microgrids with the addition of an extra Kuramoto oscillator corresponding to node 4. Using similar arguments to those exploited to get \eqref{eq:synch_freq_i}, it follows that the synchronization frequency of each new cyberlayer must fulfill the equation
\begin{equation}\label{eq:synch_freq_il}
    \bar \omega_{l}^i = \frac{P(l)+P_i}{|\mathcal{M}_l(0)|+1}, \ \ l=1,\dots,n_{\mu}.
\end{equation}
with $l$ denoting the cyberlayer the equation is referring to, and $i=4$.

\item Node 4 can now estimate the power imbalance and the cardinality of each initial island $\mathcal{M}_i(0)$, $i=1,2$, it is neighboring with. Specifically, taking $   \mathcal{M}_1(0)$ as a representative example (the derivation is the same for $\mathcal{M}_2(0)$), node 4 can locally solve, for the unknowns $\vert \mathcal{M}_{1}(0)\vert$,   $P(1)$, the equations
\begin{equation}\label{eq:synch_freq_il}
    \bar \omega_{1} = \frac{P(1)}{|\mathcal{M}_{1}(0)|}
\end{equation}
\begin{equation}\label{eq:synch_freq_il1}
    \bar \omega_{1}^4 = \frac{P(1)+P_4}{|\mathcal{M}_{1}^4|+1}.
\end{equation}
Hence node 4 can estimate through local computations the power imbalance of island $\mathcal{M}_1(0)$ and similarly that of $\mathcal{M}_2(0)$.

\item Now, if node 4 is a load, it will join the island with the largest power imbalance.
If instead node 4 is a distributed generator, it will join the island with the smallest power imbalance.
The islands are then updated with the new configurations,
e.g. see Figure \ref{fig:representation_decentralized}c. 
Note that if node 4 is a load and no neighboring microgrid has a positive power imbalance then node 4 will wait until one of them becomes positive by the addition of more generators to its neigboring microgrids. 

The procedure is iterated by all the network nodes until there are no more isolated nodes.

\end{enumerate}

\begin{remark}
Note that at every step, the number of cyberlayers that a generic node has to evaluate in order to make a decision is equal to the $$\sum_{j=1}^{n_{\mu}}\left|\mathcal{N}(\mathcal{M}_j)\cap \left( \mathcal{V} \setminus \cup_{i=1}^{n_\mu} \mathcal{M}_i\right)\right|$$
and is thus bounded by $n_{\mu} +n_{\mu}(n-n_{\mu})$.
\end{remark}

The decentralised strategy described above by means of a schematic example is generalized in Algorithm \ref{alg:Decentralized_Alg_Dentralized_Impl} in the form of a pseudocode.

\begin{algorithm}[t]
\SetKwInOut{Input}{input}\SetKwInOut{Output}{output}
 \Input{node $j^*$, neighbors $\mathcal{N}_{j^*}$}
 \Output{the microgrid index $l^*$ node $j^*$ will attach to}
  \BlankLine
 $l^* \leftarrow \emptyset$\;
 \While{$\mathcal{N}_{j^*} \cap \left(\bigcup_{l} \mathcal{M}_l \right) = \emptyset$}{ 
 idle\;
 }
 \nl \While{$l^* = \emptyset$}{ 
 $\mu_{j^*} \leftarrow \{ l \mid \exists l : k \in \mathcal{M}_l, \forall k \in \mathcal{N}_{j^*} \}$\;
   \For{$l \in \mu(j^*)$}{
        \If{$\Gamma_l^{j^*}$ does not exist}{
        create $\Gamma_l^{j^*}$\;
    }
  }
 Simulate the dynamics on all active layers\;
 \For{$l \in \mu_{j^*}$}{
 $t_l^{j^*} \leftarrow \ \lbrace \bar t: |\dot \theta_i(t)-\dot \theta_{j^*}(t)|\leq \epsilon \   \forall i \in \mathcal{M}_l \cap\mathcal{N}_{j^*}, \ \forall t\geq \bar t \rbrace$\;
 }
 $t^{j^*} \leftarrow \max_{l \in \mu_{j^*}} t_l^{j^*}$\;
 \For {$l \in \mu_{j^*}$}{
    $P(l) \leftarrow \bar \omega_l(\bar \omega_l^{j^*}-P_{j^*})/(\bar \omega_l - \bar \omega_l^{j^*})
    $\;
  }
 \If{ $\exists l : \mathcal{N}_{j^*} \subseteq \mathcal{M}_l$}
 { $l^* \leftarrow l : \mathcal{N}_{j^*} \subseteq \mathcal{M}_l$
 }
 \Else{ 
 \If{$P_{j^*} < 0$}{
  \If{$\max_{i \in \mu_{j^*}} P(i) > 0$}{
  $l^* \leftarrow \arg \max_{l \in \mu_{j^*}} P(l)$\;
  $\tilde{\omega} \leftarrow \bar{\omega}_{l^*}$\;
  }
  \Else{
  do nothing\;}
  }
  \Else{
  $l^* \leftarrow \arg \min_l P(l)$\;
  $\tilde{\omega} \leftarrow \bar{\omega}_{l^*}$\;
  }
  }
 }
 \While{$t < t^{j^*}$}{
    \For{$l \in \mu_{j^*}$}{
    \If {$\lvert \bar{\omega}_{l^*}(t) - \tilde{\omega} \rvert \ge \epsilon$}{
    go to 1\;
    }
 }
 }
  $\mathcal{M}_{l^*} \leftarrow  \mathcal{M}_{l^*} \cup j^* $\;
  $l^* \leftarrow \emptyset$\;
\caption{Decentralized partitioning}
\label{alg:Decentralized_Alg_Dentralized_Impl}
\end{algorithm}

\section{Numerical Validation}\label{sec:numerical}

In this section, we validate our partitioning algorithms on the IEEE 118 test case, for which other partitions are available in the literature. To quantitatively assess the performance of our algorithms, we will use the metrics presented in the next section.

\subsection{Partition Metrics}

We consider
%To measure the performance of a partition we make use of 
the following metrics from \cite{haddadian2017multi,Fan2021Controlled,arefifar2012supply,arefifar2013optimum}:
\begin{subequations}\label{eq:metrics}
\begin{align}
 \label{eq:adequacy}
&J_{1} \coloneqq \frac{1}{n_{\mu}} \sum ^{n_{\mu }}_{l = 1}| P(l) |,\\
\label{eq:V_dev}
&J_{2} \coloneqq \frac{1}{n_{\mu }}\sum ^{n_{\mu }}_{l =1}\left( 1-\frac{V_{\text{min},l }}{V_{\text{max},l }}\right),\\
\label{eq:Losses}
&J_{3} \coloneqq \sum_{( i,j) \in \mathcal{E}_l \ \forall l} |P_{\text{loss},ij} |,\\
\label{eq:P_dis}
&J_{4} \coloneqq \sum _{ i\in V_{k} , j\in V_{h},
k\neq h}\frac{|P_{ij} |+|P_{ji} |}{2}.
\end{align}
\end{subequations}
In \eqref{eq:metrics}:
$P(l)$ is the power imbalance in the $l$-th island and is defined in \eqref{eq:excess};
$V_{\min,l}$ and $V_{\max,l}$ are the minimum and maximum bus voltages among the nodes in the $l$-th island, respectively;
$P_{\text{loss},ij}$ is the power loss on line $(i,j)$%
\footnote{note that only the losses relative to lines that are not cut by the partition are considered.};
finally $P_{ij}$ is the power flowing on line $(i,j)$.
Hence, 
\begin{itemize}
    \item $J_1$ is the average \emph{power imbalance} in each island;
    \item $J_2$ is the average \emph{voltage deviation} in each island and is a measure of the quality of the dispatched power;
    \item $J_3$ is the sum of the \emph{power losses} in the islands;
    \item $J_4$ is the \emph{power disruption} induced by the partition, i.e., the sum of the power that was transmitted when the grid was connected on the lines cut to obtain the partition. 
\end{itemize}
We obtain all the quantities required to compute the metrics in \eqref{eq:metrics} by solving for each test case an AC Optimal Power Flow (OPF) using \textsc{Matpower 6.0} \cite{zimmerman2010matpower}.
We leave all the solver options at their default values, except for the cost function, that is chosen to be the total generated power 
$$P_{\mathrm{gen}} := \sum_{i \in \mathcal{V}^{\text{gen}}} P_{i} .$$
Note that all the input parameters required to solve the OPF are available in \textsc{Matpower} for our test case.

\subsection{IEEE test case 118}

We consider the problem of partitioning the IEEE 118 bus system in $n_{\mu}=2$ islands.
We restrict the set of generators $\mathcal{V}^{\text{gen}}$ to be composed only of the 19 synchronous machines present in this test case.
Namely, we do not consider the reactive compensators and the synchronous motors, so as to make our partition comparable with that in \cite{Fan2021Controlled}, which we will use as a benchmark.
Moreover, we assume that line 14-15 disconnects due to a three phase solid ground fault at bus 15. 

To obtain our partitions, we start by applying the procedure outlined in section \ref{sec:synch_for_sol}. Namely, we initialise the islands to the two sets of coherent synchronous machines which we take from \cite{Fan2021Controlled}. As the subgraphs of $\mathcal{G}$ defined by this initialisation are not connected, we complement the coherent generators in each set with nodes taken from the shortest paths connecting them, thus obtaining
\begin{align*}
    \mathcal{M}_1 &= \lbrace 3,5,8,9,10,12,17,25,26,30,31\rbrace,\\
    \mathcal{M}_2 &= \lbrace 45,46,49,54,59,61,65,66,69,77,80,82,\\
    &\phantom{={}} 83,85,86,87,89,98,100,103,110,111\rbrace
\end{align*}
as the node sets of the initial connected islands. 

\begin{figure}[h!]
\begin{subfigure}{0.5\textwidth}
  \centering
    \includegraphics[scale=0.5, trim = {0cm 3cm 0cm 2cm}]{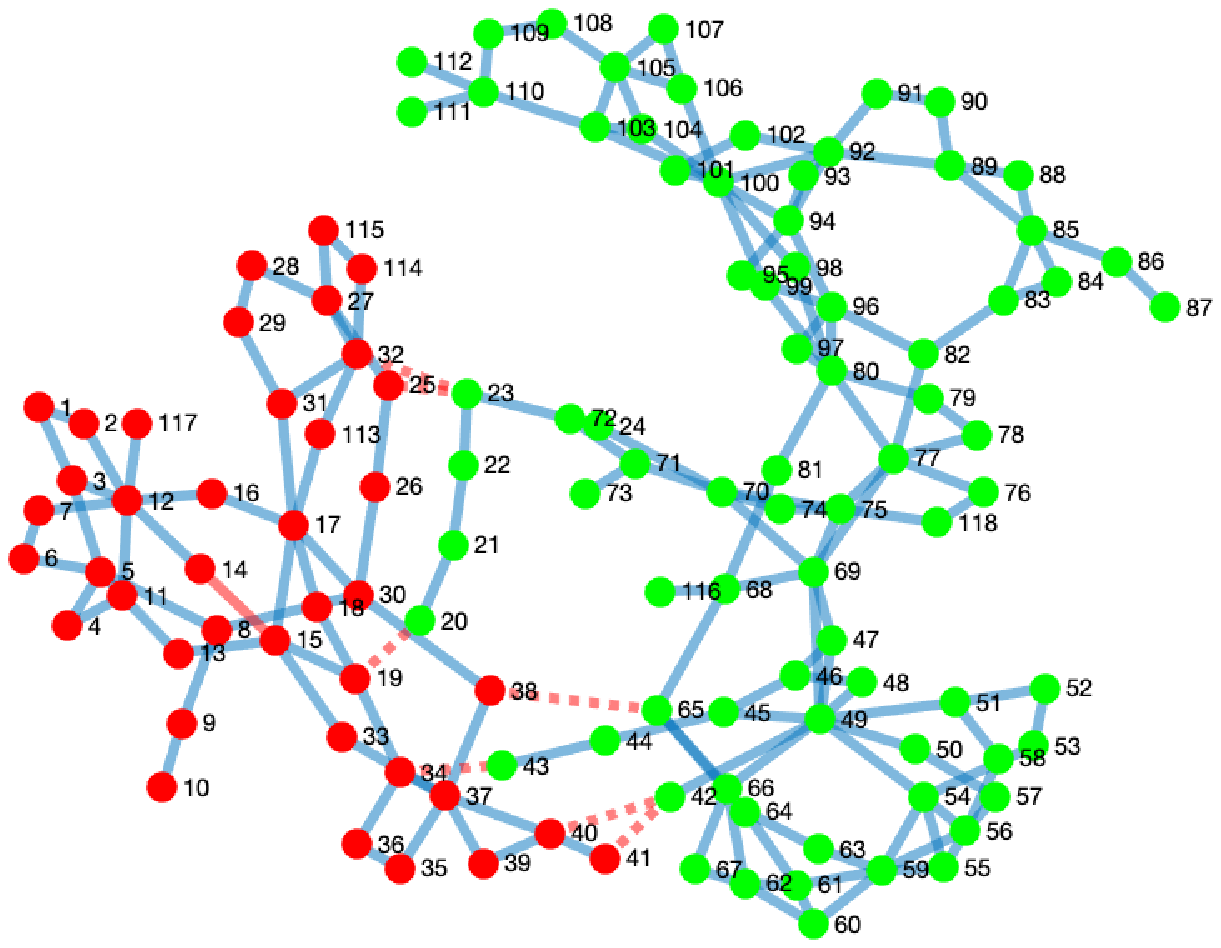}
    \caption{}
    \label{fig:Centralized_Partition_Case_118_centralized}
\end{subfigure}
\begin{subfigure}{0.5\textwidth}
  \centering
    \includegraphics[scale=0.5, trim = {0cm 3cm 0cm 0cm}]{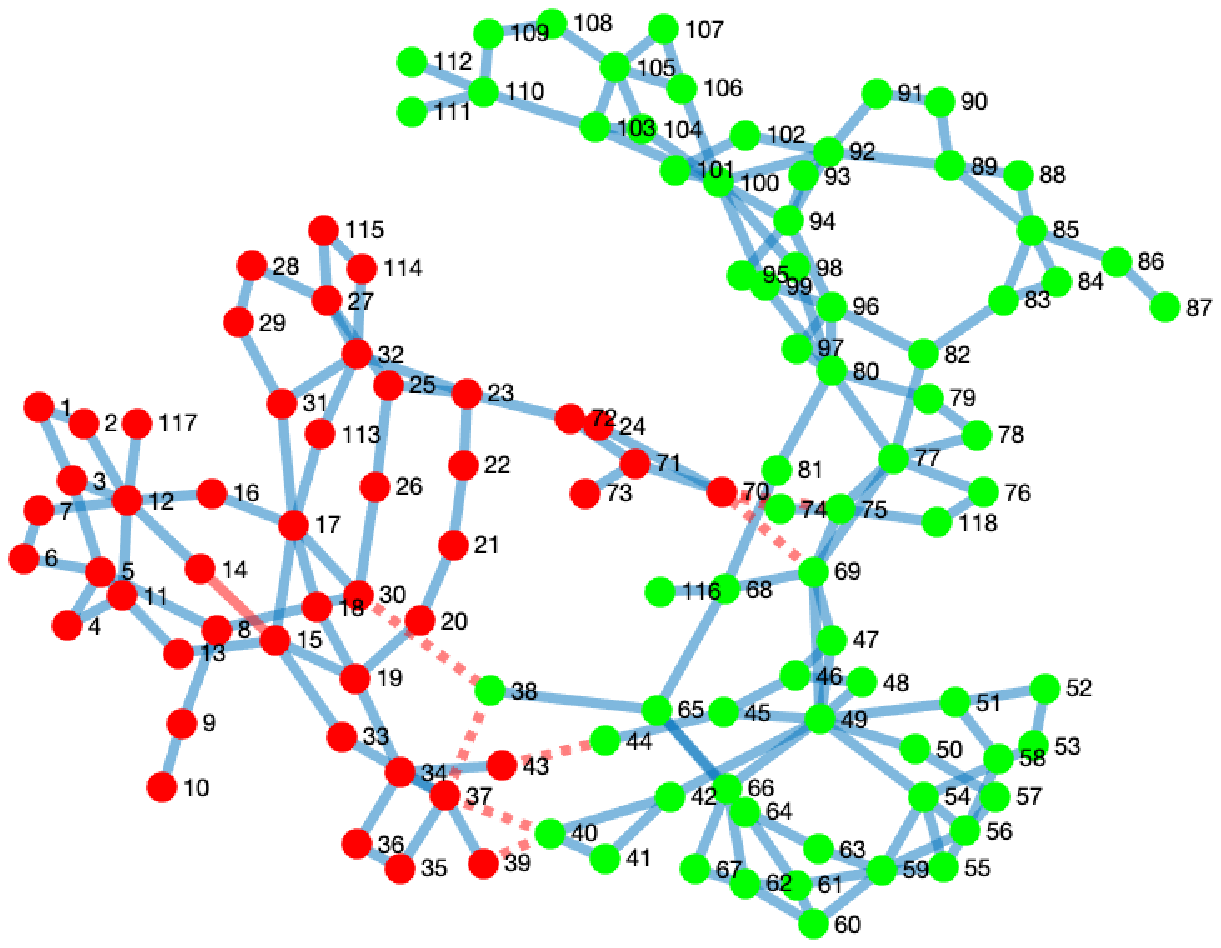}
    \caption{}
    \label{fig:Centralized_Partition_Case_118_decentralized}%[1ex]
\begin{subfigure}{0.5\textwidth}
  \centering
    \includegraphics[scale=0.5, trim = {4.5cm 3cm 0cm 0cm}]{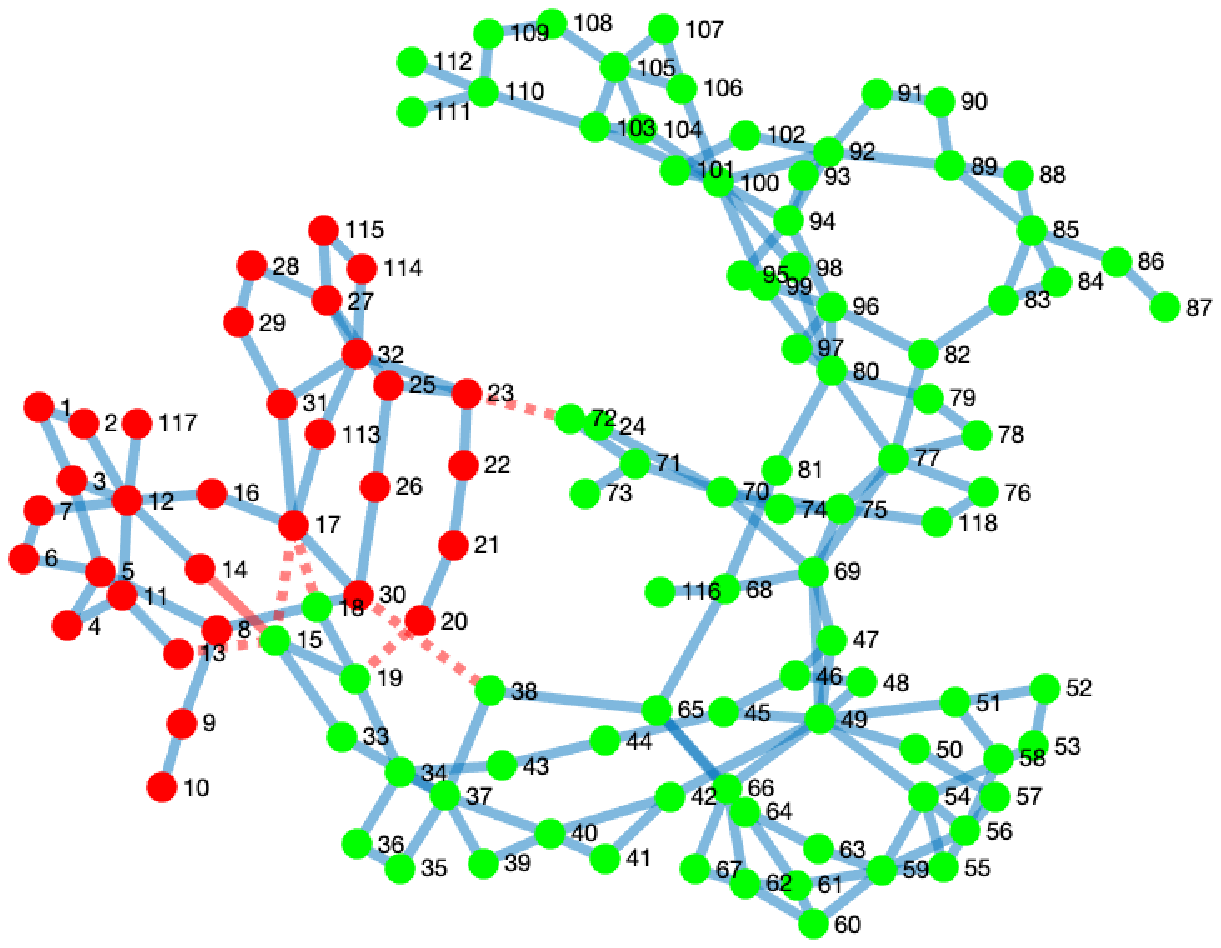}
    \caption{}
    \label{fig:Comparison_Partition_Case_118_benchmark}
\end{subfigure}
\end{subfigure}
\caption{Three different partitions of the IEEE test case 118 in two islands obtained by applying (a) Algorithm \ref{alg:Centralized_Alg}, (b) Algorithm \ref{alg:Decentralized_Alg_Dentralized_Impl}, and  (c) the procedure in \cite{Fan2021Controlled}. The nodes in red belong to island $1$, while the nodes in green belong to island $2$. Red dashed lines represent the cut-set edges dividing the islands while the failed transmission line is the red solid line.}
\label{fig:Partitions_118}
\end{figure}

Next, we apply Algorithm \ref{alg:Centralized_Alg} to obtain a partition in a centralised fashion.
The resulting islands, whose power imbalance is $P(1) = -93$ MW and $P(2) = 172$ MW, respectively, are shown in Figure \ref{fig:Centralized_Partition_Case_118_centralized}.
On the other hand, applying the decentralised Algorithm \ref{alg:Decentralized_Alg_Dentralized_Impl}, we obtain the islands in Figure \ref{fig:Centralized_Partition_Case_118_decentralized}.
In this case, the power imbalance in each island is $P(1) = -154$ MW and $P(2) = 233$ MW.

To evaluate the performance of our algorithms, we compare the values of the metrics \eqref{eq:metrics} evaluated on the partitions we obtained and on the benchmark partition in \cite{Fan2021Controlled}, also shown in Figure \ref{fig:Comparison_Partition_Case_118_benchmark}.
Let us note that in \cite{Fan2021Controlled} the authors allow for high voltage DC connections between the islands. 
As we do not consider DC transmission lines, we compute the metrics \eqref{eq:metrics} for the partition in \cite{Fan2021Controlled} neglecting their effect. Interestingly, both the partitions obtained from Algorithm \ref{alg:Centralized_Alg} and Algorithm \ref{alg:Decentralized_Alg_Dentralized_Impl} perform better than that of ref. \cite{Fan2021Controlled} in terms of power imbalance $(J_1)$. On the other hand while we observe worse values of the power disruption $(J_4)$. All three partitions exhibit instead comparable values in terms of voltage deviations $J_2$ and power losses $J_3$. Overall, both our partitions present comparable values of the metrics in \eqref{eq:metrics} with respect to that in ref. \cite{Fan2021Controlled}.

\begin{table}[h!]
    \centering
    \def\arraystretch{1}
\begin{tabular}{llll}
\hline
\textbf{Metric}                                                     & \textbf{ \cite{Fan2021Controlled} (no DC lines)}     & \textbf{Centralised}    & \textbf{Decentralised}\\ \hline
$J_1$       & $289$                    & $133$                          & $194$                                 \\ 
$J_2$       & $0.0835$                    & $0.0965$                          & $0.0944$                                 \\ 
$J_3$       & $37.7$                    & $36.4$                          & $36.8$                                 \\ 
$J_4$       & $305$                    & $384$                           & $509$                                 \\ \hline
\end{tabular}
\caption{values of the metrics computed for the partitions depicted in Figure \ref{fig:Partitions_118}.
}
   \label{tab:Partition_Metrics}
\end{table}

%\subsection{IEEE test case 300}
%\fli{This subsection will resemble closesly the one on the IEEE 118 test case. Instead of comparing the results with Emanuele's partition, we here aim at running a GA and comparing our solutions to the optimal (or near optimal) partitions).}
%\section{Conclusion}
%
%In this letter, we have coped with the problem of partitioning a power grid in a set of microgrids, or islands. This problem, arising both in the design of future Advanced Grids and in the recovery from extreme failures, is generally turned into a combinatorial optimization problem solved in centralized fashion. Here, we have taken a different route, and have exploited synchronization to design a centralised and a decentralised strategy that allow to obtain a partition in a computationally efficient fashion. We have tested our strategies by applying them to the IEEE 118 node test case for which alternative partitions are available in the literature. 

Our results suggest that the quality of our partitions, measured according to the established metrics available in the literature, is comparable to that of the partitions obtained by solving a more cumbersome combinatorial optimization problem. FUture work will be aimed at refining our methodology proving its convergence towards a viable partition of a desired grid.

\bibliographystyle{IEEEtran}
\bibliography{bibliography}%

%\bibliography{BibliographyDoc.bib}{}
%

\end{document}